\documentclass[10pt, conference, compsocconf]{IEEEtran}
\usepackage{color}

\usepackage{cite}

\ifCLASSINFOpdf
   \usepackage[pdftex]{graphicx}
\else
\fi
\usepackage{amsthm} 

\newtheorem{definition}{Definition}

\usepackage{adjustbox}
\usepackage{amsmath}
\usepackage{mathtools}
\usepackage{amssymb} 
\usepackage{enumitem} 
 \usepackage{amssymb,amsmath,caption}
 \usepackage[hang,flushmargin]{footmisc}
\makeatletter
\newcommand{\algorithmfootnote}[2][\footnotesize]{%
  \let\old@algocf@finish\@algocf@finish
  \def\@algocf@finish{\old@algocf@finish
    \leavevmode\rlap{\begin{minipage}{\linewidth}
    #1#2
    \end{minipage}}%
  }%
}
\makeatother
\usepackage[linesnumbered,ruled,vlined]{algorithm2e}
\usepackage{algorithmicx}
\usepackage{pgfplots}
\usepackage{tikz}
\usepackage{comment}
\usepackage{cite}
\usepackage{eucal}
\usepackage{url}

\begin{document}

\title{SPATO: A Student Project Allocation Based Task Offloading in IoT-Fog Systems}
\author{
   \IEEEauthorblockN{Chittaranjan Swain, Manmath Narayan Sahoo and Anurag Satpathy}
   \IEEEauthorblockA{Department of Computer Science and Engineering \\
     National Institute of Technology, Rourkela, India.
   }
    \{chittaranjanswain518, anurag.satpathy\}@gmail.com, sahoom@nitrkl.ac.in
}

\maketitle
\begin{abstract}
The Internet of Things (IoT) devices are highly reliant on cloud systems to meet their storage and computational demands. 
However, due to the remote location of cloud servers, IoT devices often suffer from intermittent Wide Area Network (WAN) latency which makes execution of delay-critical IoT applications inconceivable. 
To overcome this, service providers (SPs) often deploy multiple fog nodes (FNs) at the network edge that helps in executing offloaded computations from IoT devices with improved user experience. As the FNs have limited resources, matching IoT services to FNs while ensuring minimum latency and energy from an end-user's perspective and maximizing revenue and tasks meeting deadlines from a SP's standpoint is challenging. Therefore in this paper, we propose a student project allocation (SPA) based efficient task offloading strategy called \textit{SPATO} that takes into account key parameters from different stakeholders.
Thorough simulation analysis shows that \textit{SPATO} is able to reduce the offloading energy and latency respectively by 29\% and 40\% and improves the revenue by 25\% with 99.3\% tasks executing within their deadline.
\end{abstract}


\section{Introduction}\label{sec:introduction}
Internet of Things (IoT) has become an indispensable aspect of everyday life owing to its wide range of applications ranging from wearable devices, smart meters, connected vehicles, smart grid, smart health, intelligent transportation systems, and many more \cite{8863944}. 
In fact, the number of IoT devices supporting different applications is estimated to exceed the $30$ billion mark by $2030$ \cite{evans2011internet}. 
Typically the IoT devices are resource-constrained and rely upon the remote cloud for computation, storage, and analytics of data. 
However, cloud data centers (DCs) are usually deployed in locations that are distant from the IoT devices thereby incurring higher response time due to intermittent WAN delays and multi-hopping. 
The next-generation IoT applications not only require the processing of a huge volume, velocity, and variety of data but also demands the quality of service (QoS), location awareness, real-time mobility support, and latency-sensitive requirements. 
These specifications render the cloud-based IoT platforms impractical for modern applications.

Fog computing (FC), on the other hand, brings the cloud services closer to the users thereby improving the responsiveness of applications. 
IoT devices garner the benefits of FC by offloading computations to the nearby FNs.
These FNs deployed by the service providers (SPs) have limited resources compared to the cloud, hence offloading IoT services with heterogeneous requirements to disparate FNs is challenging. 
Considering full offloading scenarios, Adhikari \textit{et al.} \cite{8931777} proposed an application offloading strategy based on accelerated particle swarm optimization (APSO) for a hierarchical fog-cloud environment that takes into account multiple quality-of-service (QoS) parameters such as cost and resource utilization. 
Alternatively, Hussein and Mousa \cite{9006805} discussed two offloading strategies based on ant-colony-optimization (ACO) and particle swarm optimization (PSO) to load balance the assignment of tasks to FNs under communication cost and response time considerations.
Some other works that have modeled task offloading as optimization problems are discussed in \cite{zhu2018folo, 8717643}. 
Although optimization approaches may guarantee sub-optimal solutions, they suffer from the following pitfalls. 
Firstly, optimization techniques are incapable of considering contrasting objectives of stakeholders that do not align well with the system-wide objectives. 
Secondly, the optimization solvers are computationally intensive and are not scalable. 
Matching theory-based solutions overcome these drawbacks and focus on reducing the response time and energy consumption \cite{8647545, 8647845, 9201504}. 
Although these approaches resolve the concerns of optimization solutions, they face the following issues, however. 
The solution approaches discussed in \cite{8647545, 8647845, 9201504} are restricted to a single SP environment. 
Moreover, Gu \textit{et al.} \cite{gu2018joint} pointed out that the direct interaction between the FNs and IoT devices may raise security concerns such as eavesdropping and data hijacking. 
As a remedy, all communications including authentication and authorization should be carried out under the supervision of SPs, and FNs are restricted to computations.
Hence, in addition to FNs and IoT devices, we introduce SP as another entity into our model \cite{7833651}.
This addition increases the complexity as the agenda of SPs have also to be taken into consideration while generating a solution to the offloading problem. 
The overall offloading problem is analogous to the student-project-allocation (SPA) model \cite{abraham2007two}.
Additionally, the SPA-based solution converges faster to a stable allocation compared to the conventional matching-based solutions \cite{7833651}. 

In this work, we propose SPA based solution approach called \textit{SPATO} for full offloading scenarios considering an additional entity, i.e., SPs in the allocation process. 
The preferences assigned to FNs by tasks generated by IoT devices are computed considering latency and energy consumption in offloading whereas SPs rank the tasks taking into account the hosting cost and deadline.
The overall contributions of the work are as follows:
\begin{itemize}
    \item We propose SPA based efficient task offloading strategy called \textit{SPATO} that aims at optimizing multiple QoS parameters such as latency, energy, and cost.
    \item As multiple parameters are involved, the preferences of the tasks are generated using analytical hierarchy process \textit{(AHP)}. We define a new concept called \textit{provider efficiency} (PE) that is used by SPs to rank the tasks. PE takes into consideration the hosting cost and deadline of the task.
    \item To evaluate \textit{SPATO}, we compare its performance with two different baselines: \textit{SMETO} \cite{8891292}, and a \textit{RANDOM} allocation strategy. 
    Simulation results state that the proposed scheme is able to reduce the offloading energy and latency by 29\% and 40\% respectively. Moreover, \textit{SPATO} is also able to maximize the overall revenue of SPs by 25\% with 99.3\% tasks executing with their specified deadlines.
\end{itemize}

The rest of the paper is organized as follows. 
Section \ref{sec:related_work} discusses the literature that we have reviewed. 
In Section \ref{sec:sys_model} we discuss the system model in detail.
Section \ref{sec:SPA_allocation_game} talks about the SPA based solution approach.
Performance analysis of \textit{SPATO} is discussed in Section \ref{sec:perfomance_Evaluation} and conclusion are drawn in Section \ref{sec:cnls}.

\section{Related Work} \label{sec:related_work}
Task offloading in a densely connected network is proven to be $\mathcal{NP}$-Hard \cite{9201504}. In view of full offloading scenarios, Adhikari \textit{et al.} \cite{8931777} proposed a particle swarm optimization (PSO) based offloading strategy to improve the QoS parameters such as cost and resource utilization.
On the other hand, Hussein and Mousa \cite{9006805} focused on load-balanced assignment of tasks to FNs using ant-colony optimization (ACO) and particle swarm optimization (PSO) based meta-heuristics. 
Some other optimization based solutions are discussed in \cite{zhu2018folo}\cite{8717643}. 
The inherent downside of optimization solutions such as its inability to consider agendas of multiple stakeholders, extensive computational requirements, and non-scalable nature does not make it a plausible solution candidate. 
In this regard, matching theory-based solution approaches have recently gained popularity owing to their ability to capture objectives of different stakeholders via preferences, ensuring the fairness of allocation using the concept of stability and its highly scalable nature. 
In the context of matching based solution approaches Abouaomar \textit{et al.} \cite{8647545} discussed a response time-oriented 
fog user assignment considering a densely connected IoT-Fog interconnection network. 
To reduce the overall energy consumption and satisfy the heterogeneous delay requirements in multi-access edge computing environments, Gu \textit{et al.} \cite{8647845} proposed a context-aware task offloading technique based on a matching game with externalities. 
To jointly optimize the system energy and the overall latency in offloading a hybrid CRITIC and TOPSIS based ranking followed by matching is presented in \cite{9201504}.

All the above approaches do not consider the presence of a third party called the SPs that often deploy and manage these FNs. 
However, the absence of SPs raises security concerns as the FNs directly interact with the IoT devices \cite{gu2018joint}. 
Moreover, in this work, we consider the presence of SPs in addition to that of IoT devices and FNs. 
As the traditional matching involves two sets of agents, we use SPA to model the task offloading problem considering multiple QoS parameters such as energy, latency, and cost. 
Next, we discuss the system model followed by SPA based allocation procedure.
\begin{figure}[htb]
\centering
\includegraphics[width=0.5\textwidth]{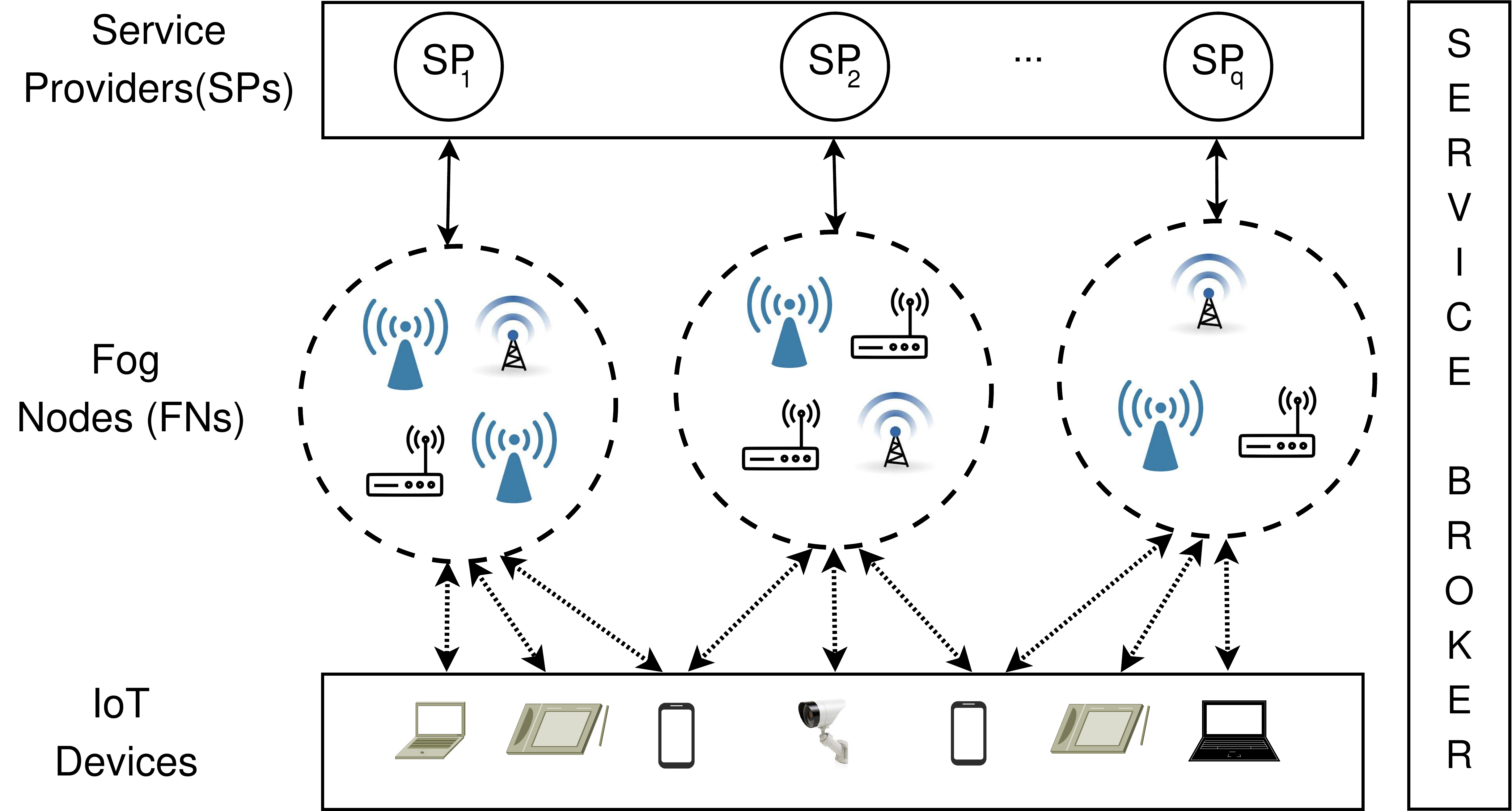}
\caption{Typical fog computing architecture with multiple SPs.}
\label{fig:fog_computing_architecture}
\end{figure}
\section{System Model} \label{sec:sys_model}
The overall architecture of an interconnected fog network is depicted in Fig. \ref {fig:fog_computing_architecture}. 
It consists of a set IoT devices denoted by $\mathbb{D}=\{d_1, d_2, d_3, \cdots, d_m\}$, where a device $d_i$ generates an atomic task $t_i \in \mathbb{T}$. 
For execution, these tasks are offloaded to FNs that are deployed and owned by a set of geo-separated SPs represented by $\mathbb{S}=\{s_1, s_2, s_3, \cdots, s_q\}$. 
Let $F_k$ be the set of FNs of SP $s_k$. 
The set of FNs in the system, $\mathbb{F} = \bigcup_{k=1}^{q} \mathbb{F}_k$.
Each FN $f_j^k \in \mathbb{F}$ corresponds to the $j^{th}$ FN of the $k^{th}$ SP. 
Additionally, we also consider the presence of a centralized service broker (SB) that performs the assignment of tasks to FNs based on a SPA matching strategy. 

For a device $d_i$, let $\mathbb{A}_i \subseteq \mathbb{F}$ be the set of feasible FNs to which $t_i$ can be offloaded. 
The offloading request corresponding to a task $t_i$ is represented using a triplet $\langle {I_{i},\, \Gamma_{i},\, \Upsilon_{i} \rangle}$, where $I_{i}$ is the size ($bits$), $\Gamma_{i}$ corresponds to computational requirement in CPU $cycles$ and $\Upsilon_{i}$ expresses the maximum tolerable delay, i.e., deadline in $sec$. 
The computational capacity of a FN $f_j^k$ is logically partitioned into a number of executable components called virtual resources units (VRUs) \cite{8647545}. 
The number of VRUs of $f_j^k$ is recorded as $C_{j}^k$, which denotes its capacity.
Each VRU of $f_j^k$ gets an equal share of the host's computational cycles, represented by $\eta_{j}^k$. 
In order to add heterogeneity to our model, we consider FNs with different capacities.
The capacity of a provider $s_k$, denoted by $C_k$, indicates the maximum number of offloaded tasks that it can serve in a given time frame using its deployed FNs $\mathbb{F}_k$ \cite{abraham2007two}. 
$C_k$ can be computed as $C_k = \sum {C_j^k}, \,\,  j \in [1, |\mathbb{F}_k|]$.
\subsection{Communication Model}
Offloading encompasses three phases: (\textit{i.}) Transmission of the task to a FN, (\textit{ii.}) Execution at the FN, \textit{(iii.)} Retrieval of computed results from the FN.
Assuming a provider operates at a bandwidth $B$ with OFDMA (orthogonal frequency division multiple access), an IoT device communicates with a FN of $s_k$ with a channel capacity $W_k \leftarrow \frac{B}{C_k}$ \cite{7517217}. 
The effective uplink rate $R_{i, j}^{k}$ from $d_i$ to $f_j^k$ is calculated as per Eq. (\ref{shannon_equ}). 
Here, $p_i$ represents the transmission power of $d_i$, $h_{i, j}^k$ is the channel gain between $d_i$ and $f_j^{k}$, and $n_0$ denotes the noise power of the channel.
\begin{equation}\label{shannon_equ}
R_{i, j}^{k} = W_k \,\, log_2 \left(1+ \dfrac{p_{i} h_{i, j}^k} {n_{0}}\right)
\end{equation}
\subsection{Parameters concerning IoTs}
The IoT devices aim at minimizing the total latency and energy consumption in offloading a task.
\subsubsection{Latency Computation}
The latency incurred in offloading a task $t_i$ to a FN $f_j^k$ is computed based on \textit{(i.)} Transmission delay to $f_j^k$, as per Eq. (\ref{eqn:transmission_time}), \textit{(ii.)} Processing delay at $f_j^k$, as per Eq. (\ref{eqn:execution time}), and \textit{(iii.)} Receiving delay from $f_j^k$, which is negligible as a small amount of processed result is to be transmitted back over a channel with comparatively high downlink rate \cite{8422316}\cite{8533343}.
\begin{equation}\label{eqn:transmission_time}
TT^{k}_{i, j}= \frac{I_i}{R_{i, j}^k}
\end{equation}
\begin{equation}\label{eqn:execution time}
ET^{k}_ {i, j}= \frac{\Gamma_i}{\eta_j^k}
\end{equation}
Thus, total latency $T^{k}_{i, j}$ in offloading is derived as per Eq. (\ref{eqn:total_ot}).
\begin{equation}\label{eqn:total_ot}
T^{k}_{i, j}= TT^{k}_{i, j} + ET^{k}_{i, j}
\end{equation}
\subsubsection{Energy Computation}
The total energy incurred in offloading a task $t_i$ form an IoT device $d_i$ to a FN $f_j^k$ consists of three components: \textit{(i.)} Transmission energy of $d_i$, as per Eq. (\ref{eqn:Energy_IoT}) \textit{(ii.)} Execution energy at $f_j^k$, as per Eq. (\ref{energy_exe}), and \textit{(iii.)} Receiving energy at $d_i$, which is ignored since it is negligible \cite{8422316}. 
In Eq. (\ref{eqn:Energy_IoT}) and (\ref{energy_exe}), $p_i$ and $p_j^{k}$ refer to the transmission power of $d_i$ and computational power of the $f_j^k$ respectively.
\begin{equation}\label{eqn:Energy_IoT}
TE^{k}_{i, j} = p_i * TT^{k}_{i, j}
\end{equation}
\begin{equation}\label{energy_exe}
EE^{k}_{i, j}= ET^{k}_{i, j} * p_j^{k} 
\end{equation}
Thus, total energy $E^{k}_{i,j}$ in offloading is computed as per Eq. (\ref{eqn:total_energy}). 
\begin{equation}\label{eqn:total_energy}
E^{k}_{i, j} = TE^k_{i, j} + EE^{k}_{i, j}
\end{equation}
The bi-objective minimization cost function for a device $d_i$ is captured by a \textit{utility score} $C_{i, j}^k$, 
which is weighted average of $T^{k}_{i, j}$ and $E_{i, j}^k$ and is computed using Eq. (\ref{eqn:utilitycost_IoT}). 
The weights $w_1$ and $w_2$ are obtained using analytical hierarchy process \textit{(AHP)} which is discussed in Section \ref{sec:ahp_based_Ranking}.
\begin{equation}{\label{eqn:utilitycost_IoT}}
C_{i, j}^k= w_1 *  T^{k}_{i, j} + w_2 * E^{k}_{i, j}
\end{equation}
\subsection{Parameter Concerning SPs}
The overall aim of a SP is to optimize its revenue by executing the maximum number of offloaded tasks within their respective deadlines.
We introduce a term called \textit{provider efficiency} (PE) that captures the efficiency of a provider taking into consideration two parameters, viz. (\textit{i.}) hosting cost at a SP, 
and (\textit{ii.}) deadline of the tasks. 
The PE $\mathcal{P}_{i}^{k}$ of a SP $s_k$ to execute a task $t_i$ can be derived as per Eq. (\ref{eqn: price}). 
The numerator expresses the overall revenue obtained by executing $t_i$ and is directly dependent on the task size $I_i$. 
However, a SP can achieve higher efficiency by executing more tasks satisfying their deadlines. 
This can only be achieved if a higher preference is assigned to tasks with closer deadlines. 
Thus $\mathcal{P}_{i}^{k}$ varies inversely with $\Upsilon_i$.
Here, ${\mathcal{C}}_k$ is a constant with unit \textit{dollar/Mbps}.
\begin{equation}\label{eqn: price}
\mathcal{P}_{i}^{k}= {\mathcal{C}}_k * \frac{I_i}{\Upsilon_i}
\end{equation}
Next, we introduce $x_{i, j}^{ k}$ to be a binary indicator variable defined as per Eq. (\ref{eqn:indicator}).
\begin{equation}\label{eqn:indicator}
x_{i, j}^{k}=
\begin{cases}
1  :\text{if $t_i$ is assigned to $f_j^k$ owned by SP $s_k$ }\\
0  :\text{otherwise}
\end{cases}
\end{equation}
The revenue obtained by a SP $s_k$, denoted as $Rev_k$, for executing a set of tasks offloaded to it is calculated as per Eq. (\ref{eqn:revenue}). 
\begin{equation}\label{eqn:revenue}
Rev_k= \sum_{i=1}^{m} x_{i,j}^k * \mathcal{P}_{i}^{k}
\end{equation}
\subsection{Problem Formulation}
The IoT devices aim at minimizing the overall latency and energy in the offloading process whereas SPs aim at maximizing the hosting cost and minimizing the number of outages due to tasks exceeding their deadlines. 
As discussed previously, PE encapsulates the dual goals of the SPs.  
The overall objective of \textit{SPATO} is presented in Eq.(\ref{eqn:objectives}).
\begin{subequations}
\begin{align}
  &\,\,\,\,\underset{\forall i\in[1, \,m]} {min}\left(C_{i, j}^k\right) \,\,\textit{and} \underset{\forall k\in[1, \,q]} {min} \left(\frac{1}{Rev_k}\right) \label {eqn:objectives}\\
 \textit{s.t.} &\qquad{} \sum_{k=1}^{q}\sum_{j=1}^{|\mathbb{F}_k|} x_{i, j}^k =1; \,\,\, i\in[1, m]\label{const_1}\\
    &\qquad{} \sum_{i=1}^{m} x_{i, j}^{k}\leq C_j^k \label{const_2};\,\,\, k\in [\,1, |\mathbb{S}|\,],\,j\in [\,1, |\mathbb{F}_k|\,] \\
  &\qquad{} \sum_{i=1}^{m}\sum_{j=1}^{|\mathbb{F}_k|} x_{i, j}^{k} \leq C_k;\,\,\, k\in [\,1, |\mathbb{S}|\,] \label{const_3} 
\end{align}
\end{subequations}
Constraint (\ref{const_1}) ensures that a task is allocated to only one FN. 
A FN can service tasks up to its capacity which is put as Constraint (\ref{const_2}) 
The total number of tasks assigned to a SP is limited to its capacity which the sum of the capacities of the FNs owned by it. This is presented as Constraint (\ref{const_3}).
The overall problem expressed in Eq. (\ref{eqn:objectives}) is a combinatorial problem and is proven to be $\mathcal{NP}$-Hard \cite{9201504}.
In fact, for a larger sample space, it is almost infeasible to solve the optimization problem in the polynomial-time frame. 
Therefore, we propose a matching theory-based heuristic to solve the task offloading problem in polynomial time.
The solution approach is detailed in the next section.
\section{Task Offloading via Student Project Allocation Game}\label{sec:SPA_allocation_game}
The SPA strategy is primarily used to assign students to projects offered by lecturers in universities \cite{abraham2007two}.
In a typical SPA setting, each lecturer has a quota indicating the maximum number of students that he can supervise.
On similar grounds, each project of a supervisor is also characterized by its quota indicating the maximum intake of the project.
Moreover, each student expresses his preference by ranking all of its acceptable projects, likewise, a lecturer constructs his preferences over students who opted for at least one of his projects. 
Motivated from \cite{abraham2007two}, we model the task offloading problem as a SPA game. 
Here IoT devices, FNs, and SPs are analogous to students, projects, and lecturers respectively. 
As preferences of IoT devices are computed based on multiple criteria, we use the analytical hierarchy process \textit{(AHP)} to obtain a unified ranking of the FNs.
\subsubsection{Ranking based on \textit{AHP}}\label{sec:ahp_based_Ranking}
The overall working of\textit{ AHP} to generate the preference profile of IoT devices is discussed subsequently. 
We highlight important steps used in the process.
A detailed discussion of \textit{AHP} can be found in \cite{9284253}.
Considering $\mathbb{C} = \{c_1, c_2, \cdots, c_{|\mathbb{C}|}\}$ to be a set of distinct criteria, we construct a pairwise comparison matrix $\mathbb{P}\in R^{|\mathbb{C}|*|\mathbb{C}|}$.
In our model $\mathbb{P}\in R^{2*2}$.
An entry $c_{r, v} \in \mathbb{P}$ denotes the relative importance of criterion $c_r$ against $c_v$.
We use a linear judgment scale to set the relative importance of criteria, due to its proven superiority \cite{9284253} over other judgment scales.
This relative importance of criteria is imposed by the stakeholder; IoT devices in our model.
After constructing the pairwise comparison matrix $\mathbb{P}$, we normalize each entry to obtain a normalized pairwise comparison matrix $\mathbb{P}'$. 
The normalized matrix is then averaged row-wise to obtain the column vector $ \mathcal{W} \in R^{|\mathbb{C}|\times 1}$ containing the weights of each criterion.
Next, the decision matrix $D_i \in R^{|A_i| \times |\mathbb{C}|}$ corresponding to each device $d_i \in \mathbb{D}$ is also normalized to $D_i^{'}$.
Finally, the global rank vector for an IoT device $d_i$, given by $\mathbb{G}_i \in R^{|A_i| \times 1}$, can be obtained 
as $\mathbb{G}_i = D_i^{'} \times \mathcal{W} $.
\subsubsection{SPA based efficient task offloading}
The analogous SPA-based matching game for task offloading can be mathematically expressed as per Definition \ref{preference_relation} and \ref{def:matchingdefination}. 
\begin{definition}\label{preference_relation}
Considering two sets of agents $\mathbb{T}$ and $\mathbb{S}$, let $P(a)$ be the preference profile of agent $a \in \mathbb{T} \cup \mathbb{S}$.
For instance, a task $t_i \in \mathbb{T}$ ranks one or more FNs from $\mathbb{F}$.
Similarly, a SP $s_k \in \mathbb{S}$ ranks some or all tasks from $\mathbb{T}$. 
\end{definition}
\begin{definition} \label{def:matchingdefination}
The matching game is based on a mapping function $\lambda:{\mathbb{T}} \cup {\mathbb{F}} \rightarrow 2^{{\mathbb{T}} \cup {\mathbb{F}}}$ such that:
\begin{subequations}
\begin{align}
\lambda (t_i) \, \subset \,\mathbb{F}  \,\,and \,\,|\lambda (t_i)|\, \leq \,1 \label{con_1}\\
\lambda (f^{k}_{j})\, \subseteq \,\mathbb{T} \,\, and\,\, |\lambda (f^{k}_{j})|\, \le \,C^{k}_{j} \label{con_2}\\
\sum_{j=1}^ {|\mathbb{F}_k|} \lambda(f^{k}_{j}) \leq C_k, \,\, k\in[1, |\mathbb{S}|] \label{con_3} \\
f^{k}_{j} \in \lambda(t_i) \Leftrightarrow t_i \in \lambda(f^{k}_{j}) \label{con_4}
\end{align}
\end{subequations}
\end{definition}
Condition (\ref{con_1}) states that a task is matched to at most one FN.
Condition (\ref{con_2}) ensures a maximum number of tasks assigned to a FN should be less than or equal to its quota.
The maximum number of tasks that can be served by a SP can be no more than $C_k$ and this is reflected in 
Condition (\ref{con_3}).
Condition (\ref{con_4}) states that a task $t_i$ is matched to a FN $f^{k}_{j}$ \textit{iff} $f^{k}_{j}$ is matched to $t_i$.
\begin{definition}\label{def:def_3}
A pair $(t_i, f_j^k)$ is a blocking pair if $f_j^k \notin \lambda(t_i)$ and the following conditions are satisfied:
\begin{enumerate}
    \item $f_j^k \in A_i$, i.e., $t_i$ finds $f_j^k$ acceptable,
    \item $\lambda(t_i)= \phi$, or $ {f_j^k} \succ_{t_i} \lambda(t_i)$, and
    \item{either}
    \begin{enumerate}
        \item $f_j^k$ is undersubscribed, i.e., $|\lambda(f_j^k)|< C_j^k$ or 
       \item $f_j^k$ is full, i.e., $|\lambda(f_j^k)| = C_j^k$ and $\exists \,\, t_{i^{'}}\in \lambda(f_j^k)$ s.t. $ {t_i} \succ_{s_k} t_{i^{'}}$ or
      \item $s_k$ is full, i.e., $|\lambda_k| = C_k$, where $\lambda_k =\cup^{|\mathbb{F}_k|}_{j=1}\lambda(f_j^k)$\,and\, ${t_i} \succ_{s_k} t_{i^{'}}$, where $t_{i^{'}}$ is the worst assigned task to $s_k$. 
    \end{enumerate}
\end{enumerate}
 \end{definition}
\begin{definition}
A matching $\lambda$ is said to be stable \textit{iff} it is not blocked by any pair of agents.
\end{definition}
The preference profile of all agents are \textit{strict} and \textit{transitive}. 
\textit{Strictness} ensures that an agent is not indifferent between any two agents of the other set implying the absence of ties. 
Considering tasks $t_x, t_y, t_z$ of $\mathbb{T}$, \textit{transitivity} implies that if an agent $f_j^k \in \mathbb{F}$ of another set has preferences of the type $t_x \succ_{f_j^k} t_y$ and $t_y \succ_{f_j^k} t_z$ then $t_x \succ_{f_j^k} t_z$ also holds.

Task $t_i$ assigns preferences to $f_j^k \in \mathbb{F}$ depending on utility score $C_{i, j}^k$ computed as per Eq. (\ref{eqn:utilitycost_IoT}).
Therefore,
\begin{equation*}
f_j^k \succ_{t_i} f_{j^{'}}^{k'} \iff  C_{i, j^{'}}^{k'}  > C_{i, j}^{k} ; \,\,\, j\neq j{'} \,\, \emph{or} \,\, k\neq k{'}
\end{equation*}
Likewise, $s_k$ assigns preferences to $t_i \in \mathbb{T}$ based on the PE $\mathcal{P}_{i}^{k}$ calculated as per Eq. (\ref{eqn: price}). Hence, 
\begin{equation*}
t_i \succ_{s_k} {t_{i^{'}}}\,\iff  \mathcal{P}_{i}^{k} >\mathcal{P}_{i^{'}}^{k}; \,\,\, i\neq i{'}
\end{equation*}

\begin{algorithm}[htb] 
\caption{SPA based efficient task offloading algorithm \textit{(SPATO)}}
\label{algorithm:algo_SPA}
 \algorithmfootnote{The Delete(x, y) operation removes x from preference list of y and vice versa.}
\KwIn{$P_i$, $ \forall t_i \in \mathbb{T}$;\,\, 
$C_k$, $P_k$, $\forall s_k \in \mathbb{S}$;\,\,
$C_j^k$, $\forall f_j^k \in \mathbb{F}$.}
\KwResult{$\lambda:{\mathbb{T}} \cup {\mathbb{F}} \rightarrow 2^{{\mathbb{T}} \cup {\mathbb{F}}}$}
\textbf{Initialize}: All $i \in \mathbb{T}$ as free and each $f_j^k \in \mathbb{F}$ and $s_k \in \mathbb{S}$ as unsubscribed.\\
 \While {$\exists \,i $ $|$  $t_i$ is free and $P_i \ne \phi$}
  {
     $f_j^k$ = most preferred FN in $P_i$ not yet proposed \\
     Send proposal to $f_j^k$ and performs provisional assignment with $t_i$ \\
    \If {$f_j^k$ is over-subscribed}
     {   
          $t_{i^{'}}$ = worst task assigned to $f_j^k$ \\
          Break the assignment between ($t_{i^{'}}$, $f_j^k$) \\
     }
     \If {$f_j^k$ is full} 
     {
          $t_{i^{'}}$ = worst task assigned to $f_j^k$ \\
        \For{each $ t_{i^*} | \,\, t_{i^{'}} \succ_{f_j^k}   t_{i^*}$ in $P_j^k$}
         {
             {
             Delete ($t_{i^{*}}$, $f_j^k$) \\
             }
        }       
     }
  \If{$s_k$ is full} 
     {
        $t_{i^{'}}$ = worst task assigned to $s_k$ \\
      \For{each $t_{i^*} | \,\, t_{i^{'}} \succ_{s_k} t_{i^*}$ in $P_k$}
         {
             Remove $t_{i^*}$ from $P_k$ \\
           \For{each $f_j^k \in \mathbb{F}_k \cap A_{i^*}$}
           {
                 {
                 Delete ($t_{i^{*}}$, $ {f_j^k}$) \\
                }
             }
         }
     }          
     
 }
\end{algorithm} 
The working of \textit{SPATO} is shown in Algorithm \ref{algorithm:algo_SPA}.
The input to the algorithm is the set of agents $\mathbb{T}$, $\mathbb{S}$ and $\mathbb{F}$; and the preferences of each agent $a \in \mathbb{T} \cup \mathbb{S}$.
The algorithm outputs a stable assignment through $\lambda$.
Initially, all IoT devices are set to be free and each FN and SP are unsubscribed.
Steps 2-4 involve each unassigned task $t_i$ sending a proposal to its most preferred FN $f_j^k$ that it has not yet proposed. 
Then we perform a provisional assignment of $t_i$ to $f_j^k$.
After performing the provisional assignment the following cases may arise. 
If $f_j^k$ is oversubscribed, then worst task $t_{i^{'}}$ assigned to $f_j^k$ is found and the provisional assignment between $t_{i^{'}}$ and $f_j^k$ is broken (Steps 5-7).
If $f_j^k$ is full, then the worst task $t_{i^{'}}$ assigned to it is identified and all tasks having preference lower than $t_{i^{'}}$ are removed from the preference list $P^k_j$ of $f_j^k$ derived form preference list of $s_k$, i.e., $P_k$ (Steps 8- 11).
Accordingly, $P_k$ is also updated.
Alternatively if $s_k$ is full, all less preferred tasks  $t_{i^*}$ than $t_{i^{'}}$ are removed from $P_k$. 
As a consequence all $f_j^k \in \mathbb{F}_k \cap A_{i^{*}}$ are also eliminated from $P_{i^*}$ (Steps 12-17).
The algorithm outputs a stable allocation with no agent having an incentive to deviate from their current allocation.
\section{Performance Evaluation}\label{sec:perfomance_Evaluation}
We have performed a simulation using the iFogSim simulator \cite{gupta2017ifogsim}. The environmental setup and analysis of the simulation results are discussed elaborately in this section.  
\subsection{Environmental Setup}
We consider an interconnected fog network with $4$ geo-separated SPs.
Each SP owns a $20$ \textit{MHz} channel which is further subdivided into $C_k$ sub-channels of equal capacity. 
The constant $\mathcal{C}_k$ for a SP $s_k$ is randomly assigned in the range [$50, 100$] \textit{dollar/Mbps}.
The IoT devices and FNs are deployed randomly over a 2-D space with coordinates generated uniformly in the range U$[0, 100]$.
The maximum coverage of IoT devices is set in the range U$[200, 500]$ \textit{m}. 
The computational capabilities of FNs are expressed in the form of VRUs which are generated following the uniform distribution U$[50, 300]$. 
The computational rate (\textit{cycles/s}) and computational power (\textit{W}) are chosen in the range U$[6, 10]$ \textit{GHz} and U$[0.35, 0.55]$ \textit{W} respectively. 
The number of IoT devices varies in the range $250$-$1000$ at an interval of $250$ per observation. 
The task specific parameters such as input size, computational demand and deadline are generated uniformly in the range U$[300, 600]$ \textit{Kb}, U$[210, 480]$ \textit{million cycles} and, U$[5, 30]$ \textit{s}, respectively. 
Considering PCS-$1900$ GSM band, the free space path loss in dB between an IoT device $d_i$ and FN $f^{k}_{j}$ is calculated as $PL_{d_i, f^{k}_{j}}=38.02 + 20 log (dist({d_i, f^{k}_{j}}))$, where $dist({d_i, f^{k}_{j}})$ is the distance between $d_i$ and $f^{k}_{j}$. The channel gain is then calculated as $h_{i,j}^{k}=10^{-\left(PL_{d_i,f^{k}_{j}}\right)/ 10}$.
The transmission power of IoT device and noise power of channels is set to $0.5$ and $10^{-10}$ \textit{W} respectively.
\subsection{Baseline Algorithms}
To assess the performance of \textit{SPATO}, we compare its behavior with two baseline algorithms, viz., (\textit{i.}) Zu \textit{et al.} \cite{8891292}, referred to as \textit{SMETO} and (\textit{ii.}) a random allocation strategy, referred to as \textit{RANDOM}. 
\textit{SMETO} is based on a one-to-many matching game aimed at reducing the energy consumption in the offloading process. 
The \textit{RANDOM} allocation strategy randomly assigns tasks to FNs. 
\subsection{Experimental Results}
\begin{figure*}
    \centering
    \begin{minipage}[t]{0.32\textwidth}
    \centering
    \includegraphics[width=\textwidth]{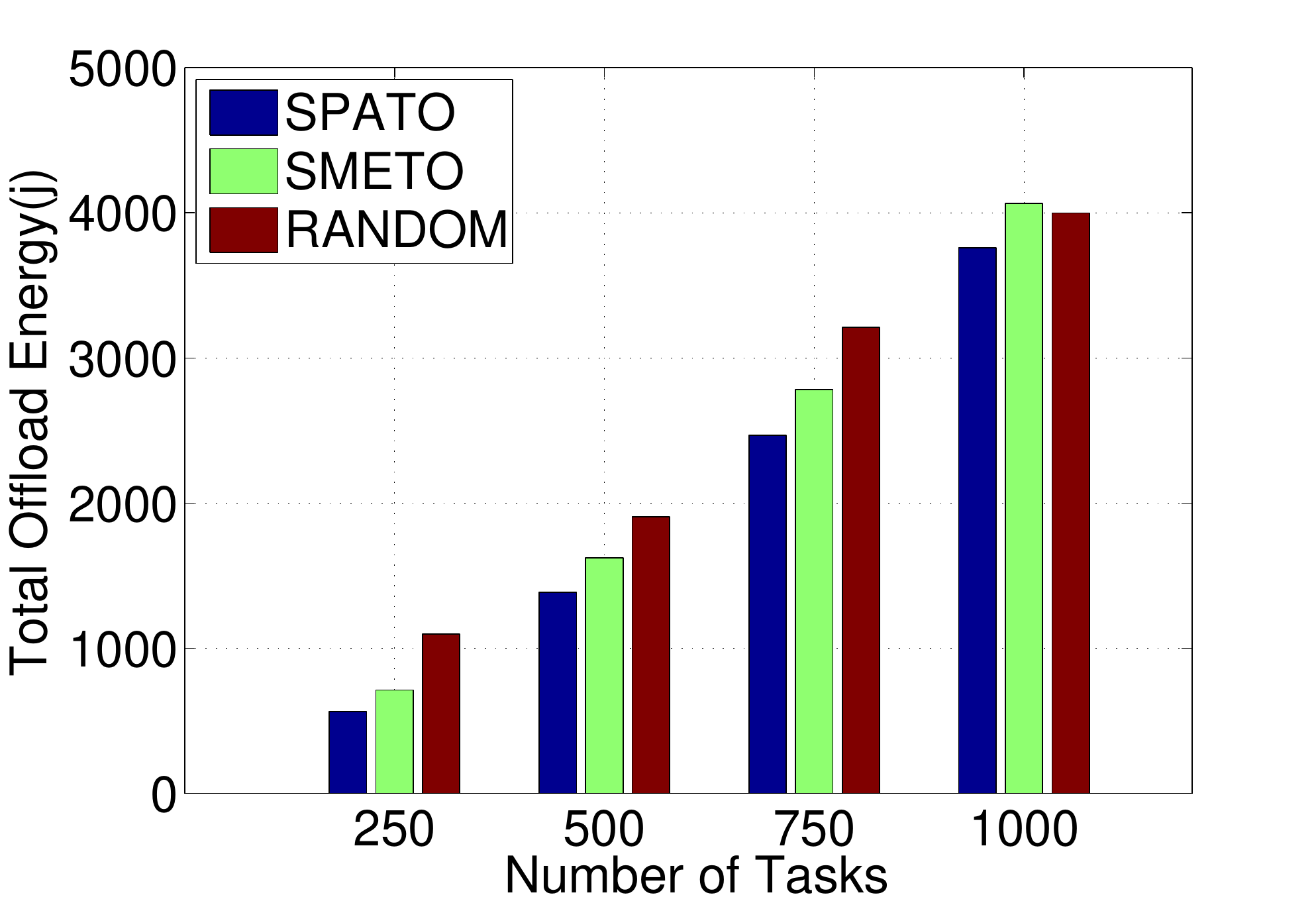}
    \caption{Total Offload Energy Vs. Number of Tasks.}
    \label{fig:systemenergy_vs_number_Of_tasks}
    \end{minipage}
    \hfill
    \begin{minipage}[t]{0.32\textwidth}
    \centering
    \includegraphics[width=\textwidth]{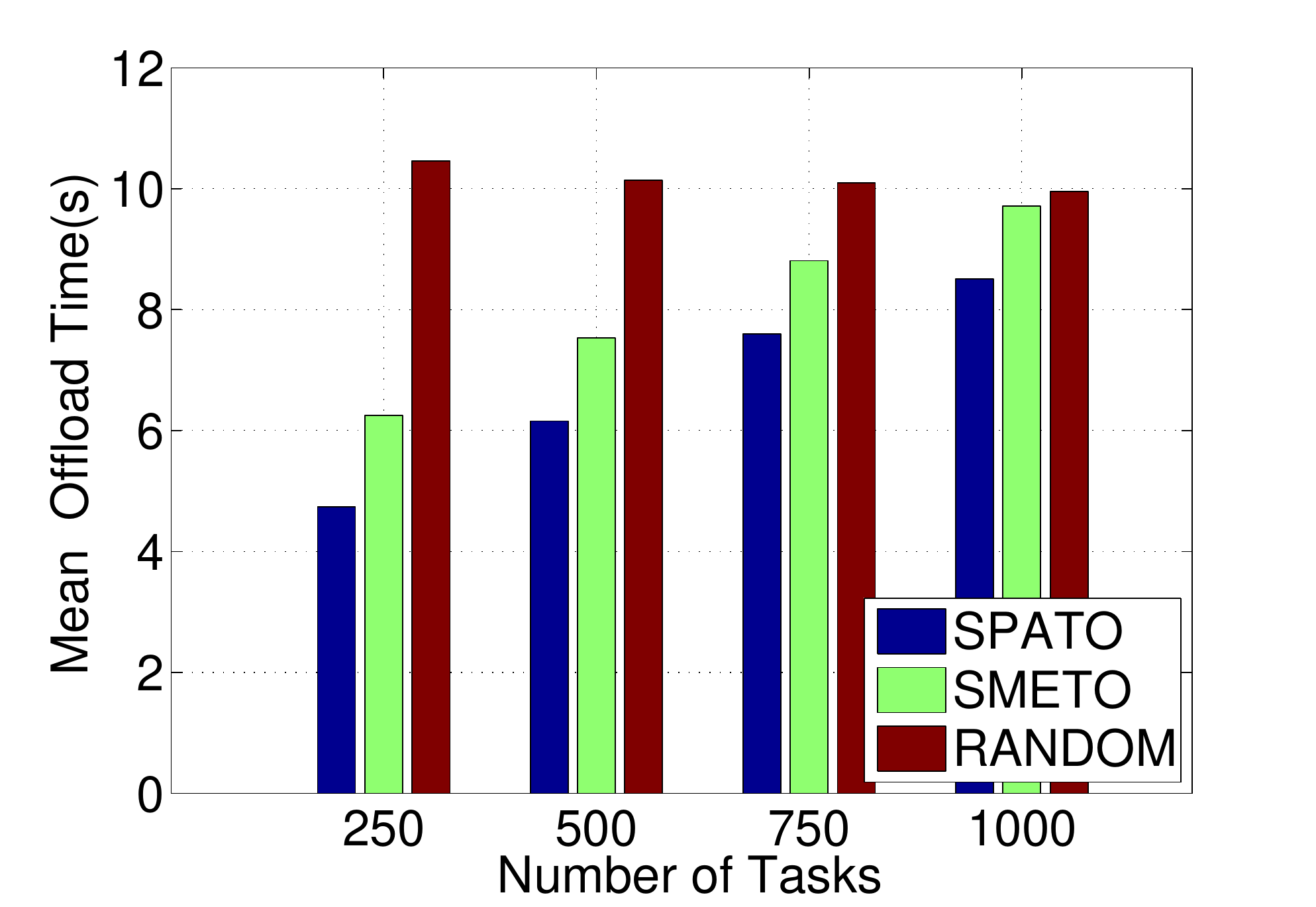}
    \caption{Mean Offload Time Vs. Number of Tasks.}
    \label{fig:completion_time_vs_number_Of_tasks}
    \end{minipage}
    \centering
    \begin{minipage}[t]{0.32\textwidth}
    \centering
    \includegraphics[width=\textwidth]{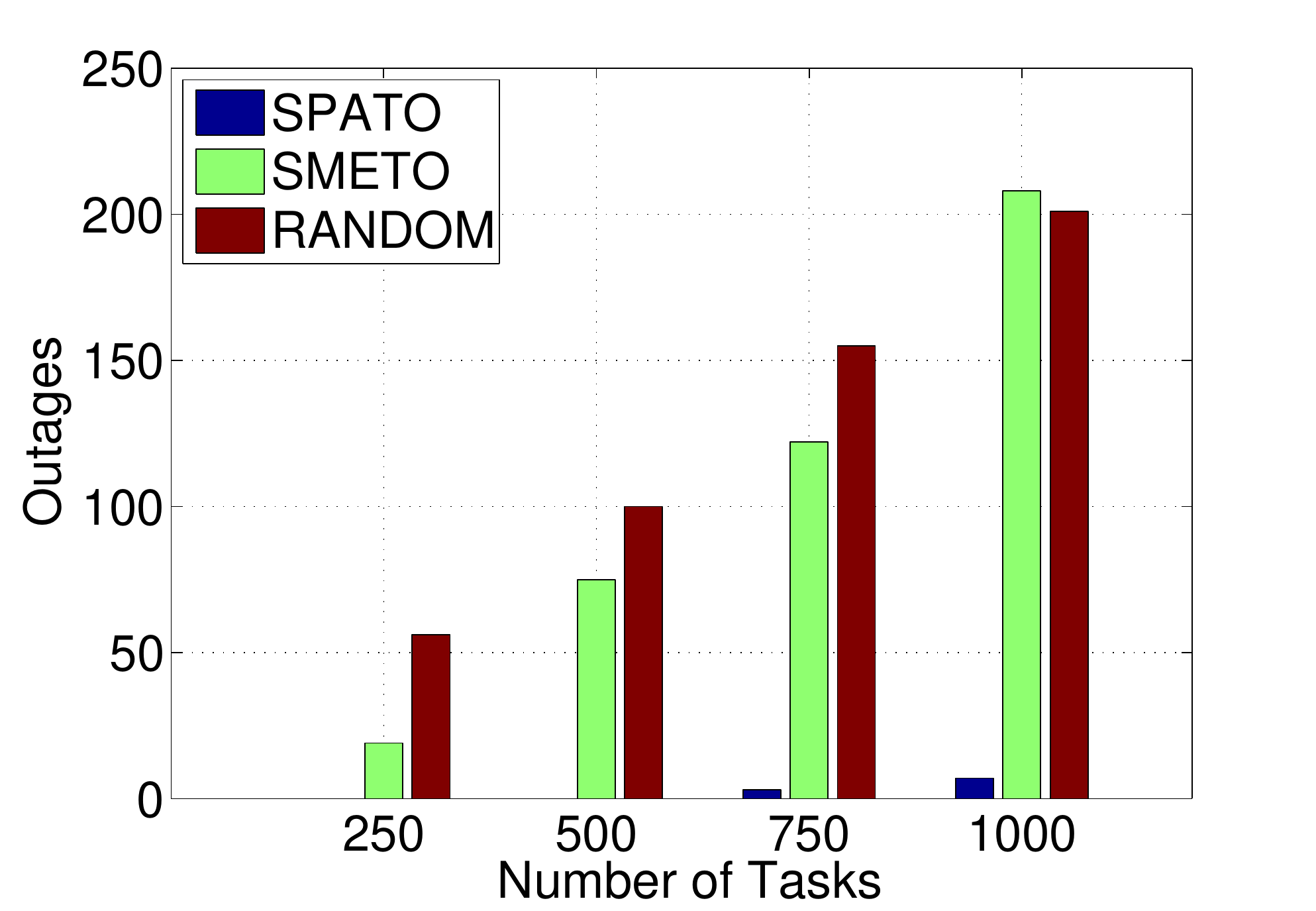}
    \caption{Outages Vs. Number of Tasks.}
    \label{fig:outages_vs_number_Of_tasks}
    \end{minipage}
\end{figure*}
Fig. \ref{fig:systemenergy_vs_number_Of_tasks} demonstrates the total energy consumed in offloading [$250$, $1000$] tasks with an interval of $250$ tasks across observations. 
As expected the offloading energy increases for an increasing number of tasks. 
The proposed algorithm outperforms \textit{SMETO} and \textit{RANDOM} baselines. 
In contrast to \textit{SMETO}, where FNs are ranked depending on transmission energy, \textit{SPATO} considers both latency and total offloading energy for ranking the FNs. 
The ranking strategy of \textit{SPATO} ensures that tasks are mostly offloaded to FNs with better computation capabilities which lead to reduced execution time, thereby reducing overall offloading time and energy. 
Fig. \ref{fig:completion_time_vs_number_Of_tasks} depicts the mean offloading time considering different number of tasks. 
The inability of \textit{SMETO} to consider offloading delay while generating preferences for IoT devices leads to elevated offloading time. 
This is because the FNs are ranked based on distance from the IoT devices which reduces the transmission time but does not ensure faster computation. 
On the contrary, \textit{SPATO} generates a unified ranking considering both energy and latency leading to reduced offloading delay.  

\begin{figure}[htb]
    \centering
    \includegraphics[width=0.32\textwidth]{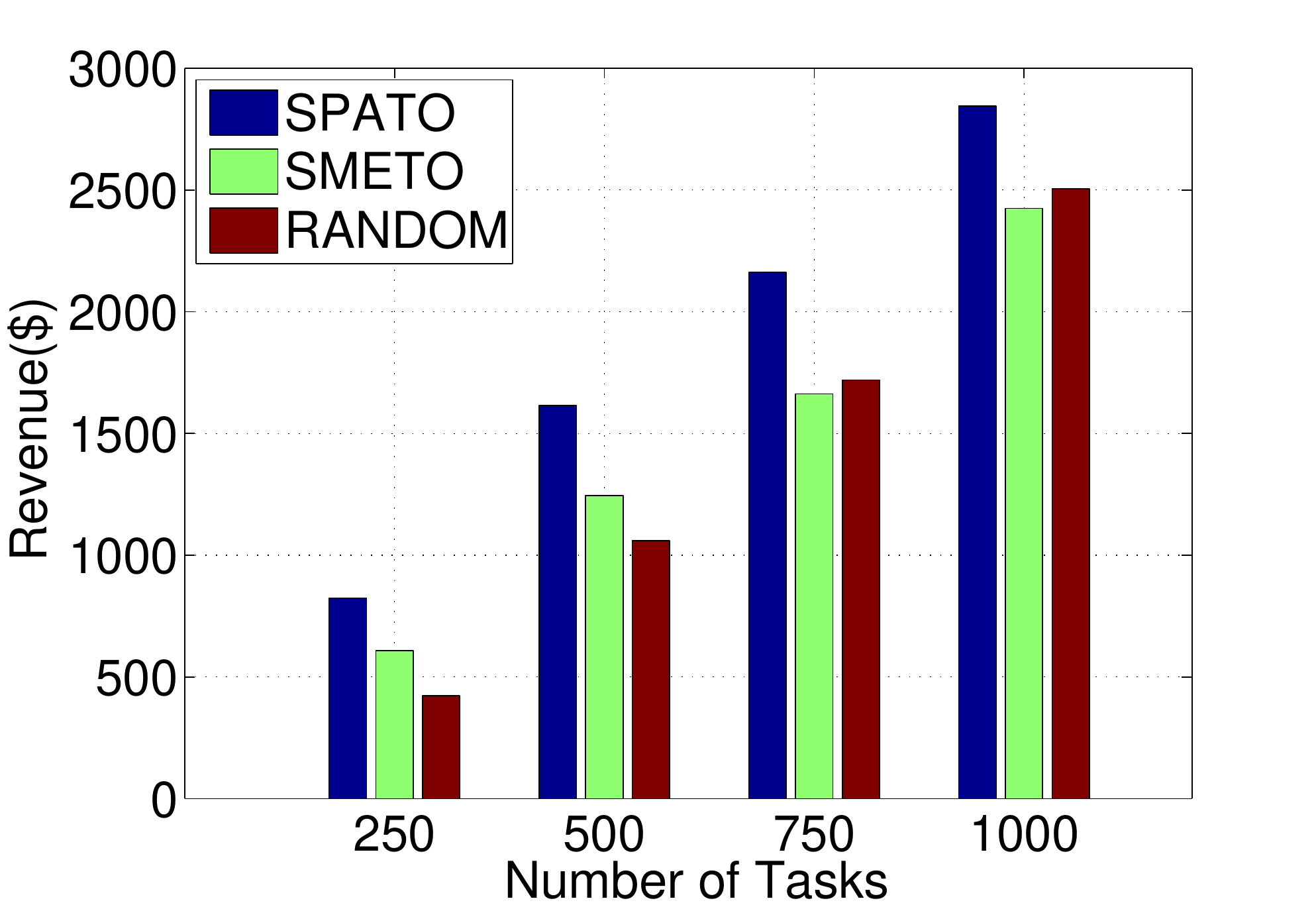}
    \caption{Revenue Vs. Number of Tasks.}
    \label{fig: Revenue_ Vs. NumberofTasks}
\end{figure}

The ranking in \textit{SPATO} considers offloading latency while generating preferences (Eq. (\ref{eqn:utilitycost_IoT})) for FNs and considers deadline while ranking the IoT devices (Eq. (\ref{eqn: price})). 
The combined effect of these rankings boosts the possibility of tasks getting executed within their specified deadlines. 
As none of the baseline algorithms consider execution time or deadline while generating preferences, they suffer from a higher number of outages which can be easily observed from Fig. \ref{fig:outages_vs_number_Of_tasks}.
The overall revenue obtained considering different approaches is shown in Fig. \ref{fig: Revenue_ Vs. NumberofTasks}. 
The proposed strategy ensures higher revenue compared to both \textit{SMETO} and \textit{RANDOM} strategies. 
The reason for this behavior is twofold.
Firstly, offloading latency considered in ranking the FNs guarantees allocation of more tasks to FNs with better computational capabilities resulting in revenue boost compared to the baseline algorithms. 
Secondly, SPs prefer large sized tasks (Eq. (\ref{eqn: price})) leading to higher revenue.
\section{Conclusion}\label{sec:cnls}
In this paper, we have proposed a model called \textit{SPATO} that aims to optimize multiple quality-of-service (QoS) parameters such as latency, energy, and cost in offloading multiple tasks in a densely connected multi-SP environment. 
Since, offloading in such a complex network is $\mathcal{NP}$-Hard, a student-project allocation (SPA) based polynomial time solution framework is developed. 
To assess the performance of the proposed technique, we compare its behavior with two baseline algorithms. 
Simulation results confirm improved performance in terms of reduced delay and energy in offloading heterogeneous tasks. 
Moreover, \textit{SPATO} is also able to maximize the SPs' revenue with minimum outages. 
As an immediate future direction to this work, we would like to consider intra and inter-channel interference arising due to channel re-usability. 
\ifCLASSOPTIONcaptionsoff
  \newpage
\fi
\bibliographystyle{IEEEtran}
\bibliography{bib_chitta}
\end{document}